\documentclass[12pt]{iopart}

\usepackage{graphicx}
\usepackage{cite}
\usepackage{bm}

\begin{document}

\title[Environment effects on the electric conductivity of the DNA]%
{Environment effects on the electric conductivity of double stranded DNA
molecules}

\author{
A.\ V.\ Malyshev\footnote{On leave from A. F. Ioffe
Physico-Technical Institute, St. Petersburg, Russia},
E.\ D\'{\i}az,
and F Dom\'{\i}nguez-Adame}
\address{GISC. Departamento de F\'{\i}sica de Materiales, Universidad
Complutense, E-28040 Madrid, Spain}
\author{V.\ A.\ Malyshev}
\address{Zernike Institute for Advanced Materials, University of  Groningen,
Nijenborgh 4, 9747 AG Groningen, The Netherlands}
\ead{a.malyshev@fis.ucm.es}

\begin{abstract}

We present a theoretical analysis of the environment effects on charge transport
in double-stranded synthetic poly(G)-poly(C) DNA molecules attached to two ideal
leads. Coupling of the DNA to the environment results in two effects:
(i)~localization of carrier functions due to the static disorder and
(ii)~phonon-induced scattering of the carrier between these localized states,
resulting in hopping conductivity. A nonlinear Pauli master equation for
populations of localized states is used to describe the hopping transport and
calculate the electric current as a function of the applied bias. We demonstrate
that, although the electronic gap in the density of states shrinks as the
disorder increases, the voltage gap in the $I-V$ characteristics becomes wider.
Simple physical explanation of this effect is provided.

\end{abstract}


\submitto{\JPCM}

\maketitle

\section{Introduction}

\label{intro}

Electronic transport through DNA molecules attached to leads still remains a
controversial topic. A number of experiments on electrical transport through
dry and wet DNA molecules revealed a variety of results. Double stranded DNA
demonstrated proximity-induced superconducting~\cite{Kasumov01},
metallic~\cite{Okahata98,Fink99,Rakitin01},
semiconducting~\cite{Porath00,Yoo01,Hwang02,Xu04,Cohen05,Roy08} and
insulating~\cite{Braun98,Storm01} behaviors. The observed differences are
often attributed to contact effects, coupling of the DNA to the
environment, and to the sequence of nucleotides. Due to the diversity of
experimental results, no consensus on mechanisms responsible for charge
transport in the DNA has been achieved so far.

Semiconducting behavior of double stranded synthetic poly(G)-poly(C) DNA was
established experimentally. The $I-V$ characteristic of the molecule revealed a
voltage gap~\cite{Porath00}. Effective Hamiltonian models, based on the
tight-binding
approximation~\cite{Iguchi97,Iguchi01,Cuniberti02,Roche03,Iguchi04,Yamada04,%
Apalkov05,Klotsa05,Rodriguez06,Malyshev07,Diaz07}, provided a reasonable
description of the semiconductor gap observed in experiments with a minimum set
of adjustable parameters.

Less effort has been devoted to study the effects of molecular vibrations on
the electric current through DNA. It was found that hopping of the charge
between the sites of guanine~(G) traps and the charge-phonon coupling
results in a staircase structure of the $I-V$
characteristics~\cite{Apalkov05bis}. The influence of vibrational modes on
the electronic properties of various types of DNA molecules (synthetic and
natural) was addressed in Ref.~\cite{Schmidt07}. It was argued that charge
transport is dominated by quasi-ballistic contributions in homogeneous DNA
and the zero-bias conductance is enhanced by the coupling to vibrations.
Dissipative effects in the electronic transport through DNA molecular wires,
comprising counter-ions and hydration shells, were investigated in
Ref.~\cite{Gutierrez05}. A bath-induced pseudo-gap opens in the
strong-coupling regime and a crossover from tunneling to phonon-assisted
transport was observed with increasing temperature. Also, it was claimed
that disorder effects smear the electronic band but have negligible impact
on the formation of the pseudo-gap.

In this work we focus on the effect of the environment on the charge transport
properties of DNA molecules. To this end, we use the tight-binding ladder
model~\cite{Iguchi97} with random base-pair energies to describe the DNA
electronic states, when disorder localizes electronic states. Charge transport
is then mediated by phonon-assisted hopping between these states, which is
described by means of a nonlinear Pauli master equation enabling us to calculate
the current as a function of the applied voltage. We study the dependence of the
$I-V$ characteristics on the magnitude of disorder.

\section{Ladder model of DNA} \label{model}

\begin{figure}[ht]
\centerline{\includegraphics[width=70mm,clip]{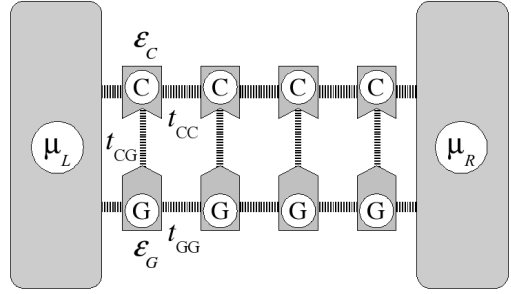}}
\caption{Schematic view of a fragment of poly(G)-poly(C) DNA molecules,
excluding the sugar-phosphate backbone, coupled to ideal leads.}
\label{fig1}
\end{figure}

We consider the ladder-like model~\cite{Iguchi97} of the poly(G)-poly(c) DNA
based on a tight-binding Hamiltonian in the nearest-neighbor approximation.
Figure~\ref{fig1} represents the schematics of the model.

\noindent The corresponding Hamiltonian reads:
\begin{eqnarray}
\left(E-\varepsilon^{C}_n\right) \psi_{n}^{C}&=&
t_{CC}\left(\psi_{n+1}^{C}+\psi_{n-1}^{C}\right)+
t_{CG}\psi_{n}^{G}\ ,\nonumber \\
\left(E-\varepsilon^{G}_n\right) \psi_{n}^{G}&=&
t_{GG}\left(\psi_{n+1}^{G}+\psi_{n-1}^{G}\right)+
t_{CG}\psi_{n}^{C}\ .
\label{amplitudes}
\end{eqnarray}
Here superscripts $G$ and $C$ label a strand, $\varepsilon^G_n$ and
$\varepsilon^C_n$ are energies of base molecules, $t_{GG}$ and $t_{CC}$ are
intra-strand transfer interactions (from now on we use $t_{CC}=t_{GG}$ for
simplicity), while $t_{CG}$ is the inter-strand interaction. In the absence
of disorder, site energies are the same along each strand.

\section{Coupling to the environment} \label{coupling}

Interactions with a random environment of solute molecules and ions surrounding
the DNA molecule can (i)~result in perturbation of the energies of base
molecules and (ii)~give rise to coupling to phonons of the bath. We account for
the former effect by considering random base energies uniformly distributed
within a box of width $\Delta$ and centered around the unperturbed energies
$\varepsilon_G$ or $\varepsilon_C$, depending on the strand. Disorder,
originated from e.g. electrostatic interactions with solute ions, can be large
and all electronic states are therefore localized at segments which are
typically shorter than the length of the DNA molecule. The system would
therefore have exponentially small transmission coefficient and direct tunneling
through the system is expected to be suppressed. On the other hand,
electron-phonon coupling can result in phonon-assisted hopping between these
localized states, i.e. incoherent charge transport.

For each realization of disorder, we diagonalize the
Hamiltonian~(\ref{amplitudes}) and calculate the scattering rate between the eigenstate
$\psi_\beta$ (with energy $E_\beta$) and another one $\psi_\alpha$ (with
energy $E_\alpha$) according to (see
Refs.~\cite{Leegwater77,Bednarz01,Shimizu01,Bednarz04,Malyshev03} for further
details):
\begin{equation}
W_{\alpha\beta}=W_0\,S(E_\beta-E_\alpha)\,
\mathcal{I}_{\alpha\beta}\;
F(E_\beta-E_\alpha,T)\ .
\label{1Wkk}
\end{equation}
Here, the constant $W_0$ stands to characterize the strength of scattering. We
assume a glassy host and take the spectral density function in the Ohmic
form $S(E_\beta-E_\alpha)=|E_\beta-E_\alpha|/t_{GG}$ widely used  in the theory
of dissipative systems (see, e.g. Ref.~\cite{Weiss}).
The temperature $T$ enters into this expression through the function
$F(E_\beta-E_\alpha,T)$ defined as
\begin{equation}
F(E_\beta-E_\alpha,T)=
\left\{
\begin{array}{lr}
1+n(E_\beta-E_\alpha), &\quad E_\beta>E_\alpha\\
n(E_\alpha-E_\beta), &\quad E_\beta<E_\alpha
\end{array}
\right.\ ,
\label{F}
\end{equation}
where $n(E_\beta-E_\alpha) = [\,\exp(|E_\beta-E_\alpha|/T) - 1]^{-1}$ is
the occupation number of the vibration mode with frequency
$|E_\beta-E_\alpha|/\hbar$. The term
\begin{equation}
\mathcal{I}_{\alpha\beta} \equiv \sum_{s=G,C}\sum_{n=1}^N
|\psi_{\alpha,n}^s|^2|\psi_{\beta,n}^s|^2
\label{overlap}
\end{equation}
represents the overlap integral of electronic probabilities for the states
$\psi_\alpha$ and $\psi_\beta$.

We describe the process of charge transport by means of the Pauli master
equation for the populations $P_\alpha$ of the eigenstates $\alpha$:
\begin{equation}
\dot{P_{\alpha}}=\Gamma_{\alpha}^{L}(f_{\alpha}^{L}-P_{\alpha})
+\Gamma_{\alpha}^{R}(f_{\alpha}^{R}-P_{\alpha})
+\sum_{\beta=1}^{2N}
\Big[(1-P_{\alpha})W_{\alpha\beta}P_{\beta}-(1-P_{\beta})W_{\beta\alpha}
P_{\alpha}\Big]\;
\label{Pauli}
\end{equation}
where $\alpha=1,2,\cdots,2N$ and $f_{\alpha}^{L,R}$ are the Fermi
distribution functions for the left and right leads:
\begin{equation}
f_{\alpha}^{L,R}=\Big[1+\exp\Big(\frac{E_\alpha-\mu_{L,R}}{T}\Big)\Big]^{-1}\ ,
\end{equation}
$\mu_{L}=E_F+eV$ and $\mu_{R}=E_F$ are the chemical potentials of the left and
right leads, $E_F$ is the Fermi energy at equilibrium taken to be in the
middle of the non-disordered DNA band gap, which is the case for Au
contacts~\cite{Xu05}. The terms
$\Gamma_{\alpha}^{L}=\gamma(|\psi_{1}^G|^2+|\psi_{1}^C|^2)$ and
$\Gamma_{\alpha}^{R}=\gamma(|\psi_{N}^G|^2+|\psi_{N}^C|^2)$ measure the
coupling between leads and the eigenstate $\alpha$, with the parameter
$\gamma$ being the strength of the coupling.

We are interested in the steady state solution of Eq.~(\ref{Pauli}). Solving
the corresponding system of nonlinear algebraic equations by an iterative
method that guarantees the condition $0\leq P_{\alpha}\leq 1$, we obtain the
stationary current as:
\begin{equation}
I(V) = \sum_{\alpha=1}^{2N}\Gamma_{\alpha}^{R}(f_{\alpha}^{R}-P_{\alpha})\ .
\label{current}
\end{equation}

\section{Results} \label{results}

In all calculations, we considered $30$ base-pair poly(G)-poly(C) DNA
molecules and used the following model parameters: unperturbed site energies
$\varepsilon^G_{n}=1.14\,$eV and
$\varepsilon^C_{n}=-1.06\,$eV~\cite{Mehrez05}, while hopping integrals were
adjusted~\cite{Malyshev07} to reproduce the current-voltage characteristics
measured in experiments on dry poly(G)-poly(C) DNA~\cite{Porath00}:
$t_{GG}=t_{CC}=0.27\,$eV and $t_{CG}=0.25\,$eV. These values are within reasonable
parameter intervals~\cite{Macia06}. We assumed that the temperature is
slightly below the freezing point of the environment ($T=273$K) which
allowed us to neglect all effect related to dynamic disorder (time-dependent
fluctuations of the configuration of solute ions surrounding the molecule)
and validates the model, namely we consider only interaction with the
phonons of the thermal bath. The parameter $\gamma$ was found to only
influence the amplitude of the current and was taken to be $\gamma=W_0$
in all calculations.

\begin{figure}[ht]
\centerline{\includegraphics[width=0.5\textwidth,clip]{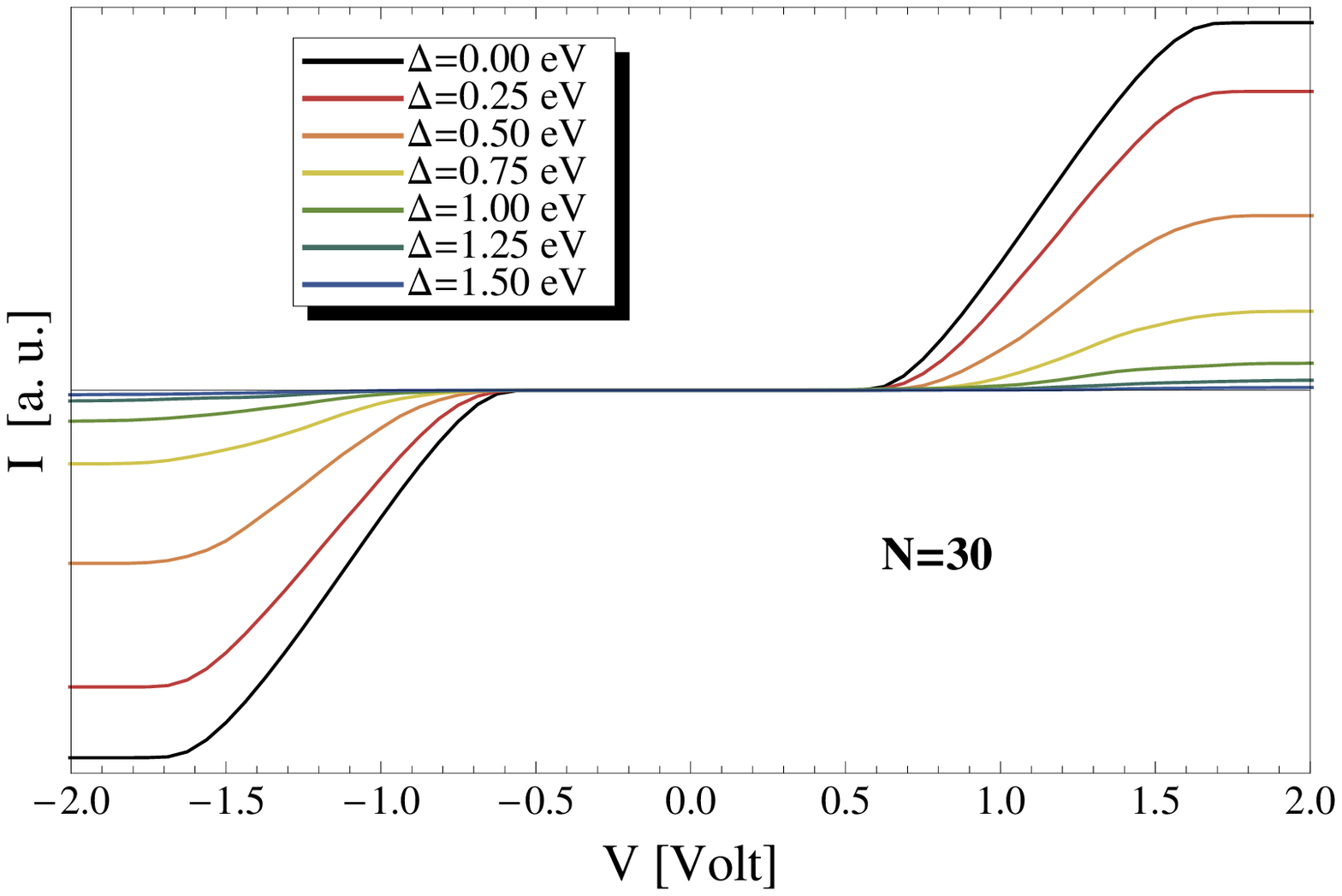}}
\vspace{2mm}
\centerline{\includegraphics[width=0.5\textwidth,clip]{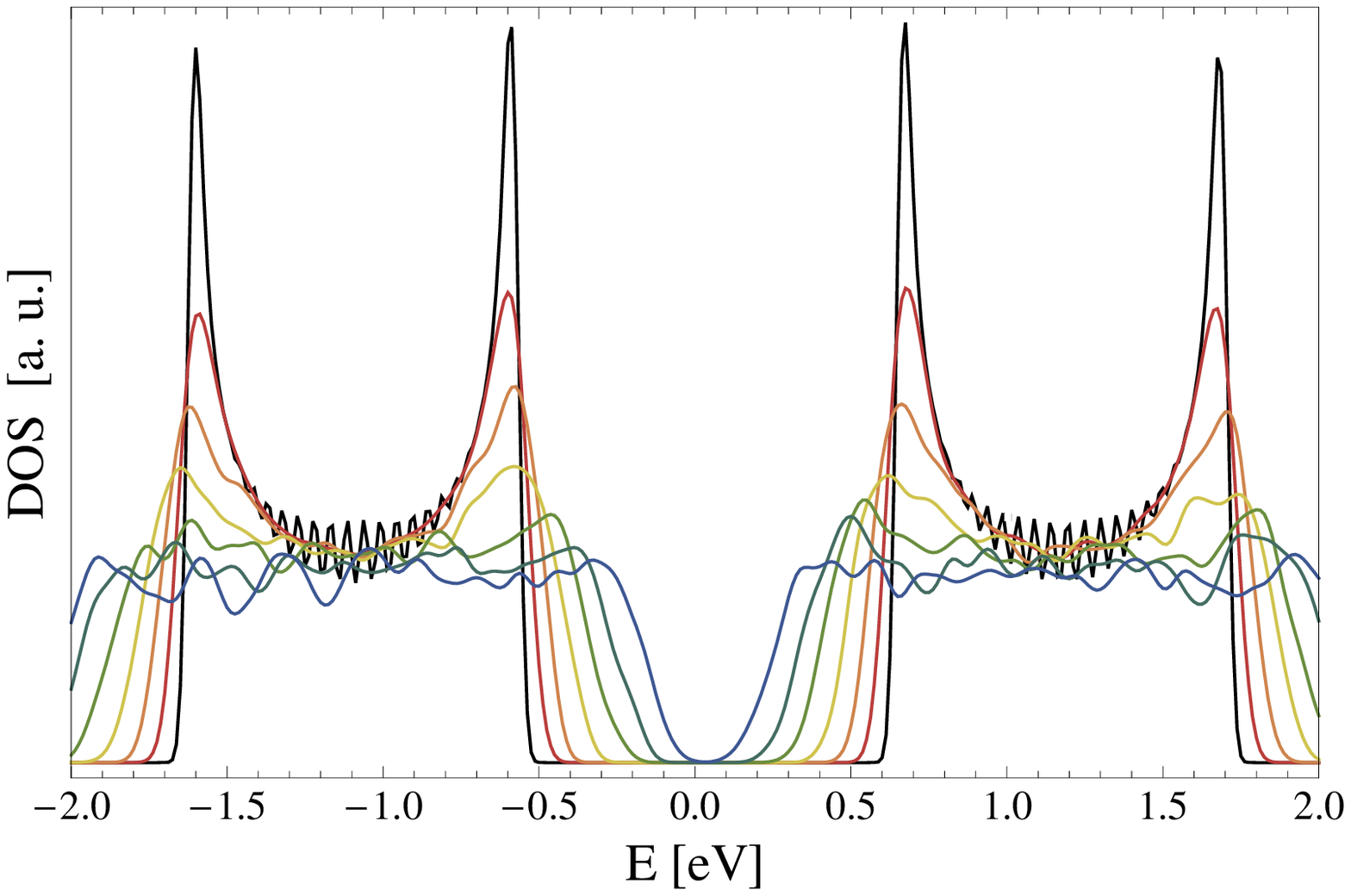}}
\caption{Current-voltage characteristics obtained for different
magnitudes of disorder indicated in the legend (upper panel) and
the corresponding density of states (lower panel).}
\label{fig2}
\end{figure}

In the upper panel of Fig.~\ref{fig2} we present the calculated
current-voltage characteristics obtained for different magnitudes of
disorder $\Delta=0\div1.5\,$eV (indicated in the legend). The lower panel
shows the corresponding density of states. These results were averaged over
$100$ realizations of disorder for each value of $\Delta$. The figure
demonstrates that the electronic gap in the density of states is shrinking
upon increasing the disorder while the voltage gap becomes wider, which
could seem counterintuitive. The closure of the electronic gap is the usual
consequence of disorder. It is due to the appearance of disorder-induced
states in the bare band gap, which form the tails of the density of states.
It may seem that these states can give rise to current at smaller voltage
drops and to the consequent shrinking of the voltage gap. Fig.~\ref{fig2}
shows just the opposite dependence of the voltage gap, namely it broadens
with increasing disorder.

In order to understand this counterintuitive dependence of the voltage gap,
the following reasoning is in place. Although there are states in the energy
gap (i.e. between bare band edges: $|E|\le 0.5$ eV in Fig.~\ref{fig2}),
these states are strongly localized, so that direct tunneling via them is
suppressed. In this case, only phonon-assisted hopping can give rise to
charge transport. If two states are localized at different segments of the
DNA, their wavefunctions have exponentially small overlap
$\mathcal{I}_{\alpha\beta}$ [see Eq.~(\ref{overlap})], and therefore, the
scattering rate $W_{\alpha\beta}$ between the two states, given by
Eq.~(\ref{1Wkk}), is small. On the other hand, for two well-overlapped
states, the typical energy spacing is large due to the quantum mechanical
level repulsion~\cite{Malyshev07bis}. Such level spacing grows on increasing
the disorder strength, reducing the thermally activated conductivity.

The typical energy separation $\delta E$ between well-overlapped states can
be estimated using the argumentation described in
Refs.~\cite{Malyshev91,Malyshev95,Malyshev01}, where it was applied to the
energy range close to the band-edge. Here, we use a similar reasoning for
the band center. Briefly, a state extended over a segment of size $N^*$
senses not the bare on-site disorder, but rather a reduced one (averaged
over the localization length) with a typical magnitude of
$\sigma/\sqrt{N^*}$, were $\sigma$ is the standard deviation of the on-site
disorder distribution ($\Delta/\sqrt{12}$ in our case). The energy
separation $\delta E$ is a function of $N^*$ (the latter magnitude is
considered to be large: $N^* \gg 1$). Then the equality $\delta E(N^*) =
\Delta/\sqrt{12N^*}$ provides a self-consistent estimate for $\delta E(N^*)$
(see Refs.~\cite{Malyshev91,Malyshev95,Malyshev01} for more details).
Applying this equality to states in the center of the band associated with
the G strand and taking into account that, for these states, $\delta E(N^*)
\approx 2\pi |t_\mathrm{GG}|/N^*$ we obtain
\begin{equation}
\delta E \approx
\frac{1}{24\pi}\> |t_\mathrm{GG} \> |\left|\frac{\Delta}{t_\mathrm{GG}}\right|^2 \ .
\label{de}
\end{equation}
For the typical considered disorder, $\Delta \approx 1$ eV the above
estimate gives $\delta E \approx 50$ meV, the energy being larger than the
considered temperature ($\approx 25$ meV). The latter means that the phonon
occupation number in (\ref{F}) is small: $n(|E_\beta-E_\alpha|/T)\ll 1$, and
suggests that phonon-assisted scattering from lower to higher states is
suppressed and almost no temperature activated transport can take place in
the system; hops from \emph{higher to lower well-overlapped states}
constitute therefore the dominant scattering process. Then, the only way for
a charge carrier to hop from one lead to another is to make a series of
cascade-like down hops over well-overlapped states. Appropriate sets of such
states can only be found in the energy region above the bare energy band
edge~\cite{Malyshev03,Malyshev06}. Disorder smears out the band edge, pushing
the boundary of this region of ``conducting'' states up towards the band
center (this boundary is analogous to the diffusion mobility
edge~\cite{Malyshev06}). Thus, for a larger value of disorder, such diffusion
mobility edge lies at a higher energy and, therefore, a greater voltage is
required to induce electric current through the system. The latter
explains the observed dependence of the voltage gap on the disorder
magnitude.

The number of
cascade states is of the order of $N/N^*$, where $N^*$ stands for the
localization length in the appropriate part of the spectrum. Because
consecutive cascade states should overlap well, they are separated by the
energy of the order of $\delta E$ given by Eq.~(\ref{de}). Then the smallest
difference between the highest and the lowest energy levels in a
cascade-like set of states can be estimated as
\begin{equation}
\Delta E_\mathrm{min} \approx \frac{N}{N^*}\,\delta E  \approx
\frac{N}{18(4\pi)^3}\> |t_\mathrm{GG}| \> \left|\frac{\Delta}{t_\mathrm{GG}}\right|^4\ .
\label{em}
\end{equation}
In deriving of Eq.~(\ref{em}) we have used $\delta E \approx
\Delta/\sqrt{12N^*}$ (see the preceding paragraph for details). Bearing in
mind that the lowest state in a cascade should be about the bare band edge
$E_0\approx 0.5$ eV, we conclude that the quantity $E_0+\Delta
E_\mathrm{min}$ shows the dependence of the voltage gap on both the disorder
and the system size. The former supports the above argument on the increase
of the gap with disorder while the latter suggests that the voltage gap
increases with the system size.

Note that the latter trend could be reversed for very short DNA chains whose
length is of the order of the smallest typical localization length. In this
situation all states are coupled to both leads and contribute to charge
transport. Thus, the current appear as soon as the Fermi level of a contact
aligns with the lowest state in the energy gap and therefore the voltage gap
coincides with the energy gap and decreases with disorder (see
Fig.~\ref{fig2}). However, in experiments the chain length is typically much
larger that the charge carrier localization length, $N \gg N^*$, so the
dependence given by Eq.~(\ref{em}) should hold, which suggests, in
particular, that long DNA chains are insulating.

\section{Conclusions}

We considered theoretically the charge transport through a synthetic
double-stranded poly(G)-poly(C) DNA molecule attached to two ideal leads and
embedded into a random static environment (e.g. solvent below its freezing
temperature). A nonlinear Pauli master equation for the populations of
localized electronic states was used to describe the hopping transport of
charge carriers. We demonstrated that the voltage gap becomes wider as
disorder increases. The calculated $I-V$ curves could indicate that the
conductivity is band-like, although the charge transport is incoherent. This
suggests that care should be taken when concluding on the nature of the
charge transport by examining only $I-V$ characteristics. The proposed
method of electric current calculation is applicable for a broader range of
systems, e.g., organic polymers.

\ack

This work was supported by Ram\'{o}n y Cajal Program, MEC (Project MOSAICO),
and BSCH-UCM (Project PR34/07-15916).

\end{document}